\documentclass[aps,pra,preprint,groupedaddress,showpacs,showkeys]{revtex4-1}
\usepackage{color}
\usepackage{graphicx}
\usepackage{amsmath,amssymb}
\usepackage{amsfonts}
\usepackage{amsmath,amssymb}

\newcommand{\chem}[1]{\mbox{{$\rm #1$}}}

\newcommand{\ds}{\displaystyle}

\newcommand{\A}{\mbox{\rm {\AA}}}

\newcommand{\ket}[1]{\mbox{$\lvert{#1}\rangle$}}
\newcommand{\bra}[1]{\mbox{$\langle{#1}\rvert$}}
\newcommand{\braket}[2]{\mbox{$\langle{#1}\rvert{#2}\rangle$}}

\newcommand{\proA}{\mbox{\rm {\A}}\mbox{$^{-1}$}~}

\def\<={\raisebox{-0.5ex}{\small $\stackrel{<}{\sim}$}}
\def\>={\raisebox{-0.5ex}{\small $\stackrel{>}{\sim}$}}
\newcommand{\Eq}[1]{Eq.~({\ref{#1}})}
\newcommand{\Fig}[1]{Figure~{\ref{#1}}}

\newcommand{\kb}{\mbox{$k_{\rm B}$}}

\newcommand{\ddxps}[1]{\frac{\ds \partial^2}{\ds {\partial #1}^2}   }

\newcommand{\dif}{\mbox{\rm d}}
\newcommand{\subi}[2]{\mbox{$ #1_{\rm #2}$}}

\newcommand{\spr}{^\prime}
\newcommand{\dpr}{^{\prime\prime}}

\newcommand{\iu}{\chem{i}}

\newcommand{\abs}[1]{\lvert #1\rvert}

\newcommand{\Exp}[1]{{\rm e}^{\ds #1}}

\newcommand{\dxsq}{\delta_x^2}
\newcommand{\dxsqi}{\breve{\delta}_x^2}
\newcommand{\rav}[2]{\left\langle #1\right\rangle_{#2}}
\newcommand{\dxsqa}{\rav{{\delta}_x^2}{v}}
\newcommand{\Tr}[2]{\chem{Tr}\left(\hat{#1}\hat{\rho}#2\right)}
\newcommand{\rhoop}{\hat{\rho}}
\newcommand{\rhotop}{\hat{\rho}^{(T)}}
\newcommand{\rhot}[1]{{\rho}^{(T)}_{#1}}
\newcommand{\Uop}{\hat{U}}
\newcommand{\Hop}{\hat{H}}

\newcommand{\psit}{\psi^{(T)}}
\newcommand{\tc}{\subi{t}{c}}

\newcommand{\tb}{\subi{t}{b}}
\newcommand{\phik}[1]{\ket{\phi_{#1}}}
\newcommand{\Dq}{\subi{D}{q}}
\newcommand{\infint}{\int\limits_{-\infty}^{\infty}}

\newcommand{\erf}{\chem{erf}}

\newcommand{\ttheta}{\tilde{\theta}}
\def\Xint#1{\mathchoice
      {\XXint\displaystyle\textstyle{#1}}%
      {\XXint\textstyle\scriptstyle{#1}}%
      {\XXint\scriptstyle\scriptscriptstyle{#1}}%
      {\XXint\scriptscriptstyle\scriptscriptstyle{#1}}%
      \!\int}
   \def\XXint#1#2#3{{\setbox0=\hbox{$#1{#2#3}{\int}$}
        \vcenter{\hbox{$#2#3$}}\kern-.55\wd0}}
   
   \def\dashint{\Xint-}
\newcommand{\fint}{\dashint}
\newcommand{\finfint}{\fint\limits_{-\infty}^{\infty}}
\newcommand{\pfinfint}{\fint\limits_{0}^{\infty}}
\newcommand{\vm}{v_T}
\newcommand{\Gs}{\subi{G}{s}}
\newcommand{\dxsqc}{\tilde{\delta}_x^2}
\newcommand{\om}{\omega}
\newcommand{\PISF}{\subi{\phi}{ISF}}
\begin{document}
\title{Mean square displacement of a free quantum particle\\ in a
  thermal state}  
\author{Roberto Marquardt}
\email[corresponding author: ]{roberto.marquardt@unistra.fr}
\affiliation{Laboratoire de Chimie Quantique - Institut de Chimie -
{UMR~7177~CNRS/Unistra}\\
Universit\'e de Strasbourg\\
4, rue Blaise Pascal - CS 90032 - 
67081 STRASBOURG CEDEX - France}
\date{\today}
\begin{abstract}
The mean square displacement $\rav{\left(x(t)-x(0)\right)^2}{}$ of the
position $x$ of a free particle of mass $m$ at thermal equilibrium is
evaluated quantum  
mechanically.
An analytical expression is obtained which shows an initial quadratic
increase of the  
mean square displacement with time and later on a linear growth, 
with the slope $\hbar/m$, quite at variance with the result from
classical statistical mechanics. 
Results are
discussed in relation  
to observables 
from helium scattering or 
spin-echo experiments, and their possible 
interpretation in terms of the classical and quantum mechanical expression
for the mean square displacement of an essentially free particle.
\end{abstract}
\maketitle
\section{Introduction}

This paper is about the mean square displacement (MSD)
$\dxsq(t) = \rav{(x(t)-x(0))^2}{\theta}$
of a thermalized, but otherwise free moving quantum particle, the position of
which at time $t$ is $x(t)$. A thermalized state can be
characterized by a set of random numbers 
$\theta$ and 
$\rav{\cdot}{\theta}$ is the
average in the sense of the arithmetic mean over these
states. 
The main motivation of the present work is the study of the
diffusion of particles within the laws of quantum mechanics. Some of the results
presented here should also be  
relevant in quantum transport theory.

In classical mechanics, a particle at thermal
equilibrium with its environment moves with statistically distributed
positions and 
velocities. For a free particle, $x(t)=x(0)+vt$ with
$\rav{x(0)}{\theta} = 0$ and $\rav{v}{\theta}=0$ but 
$\dxsq(t) = \rav{\left(v\,t\right)^2}{\theta} = \rav{v^2}{\theta}\,t^2$, and
$\rav{v^2}{\theta} = \kb T\,/\,m$ for a thermal ensemble.
Here $T$ is the temperature and 
$\kb$ is the Boltzmann constant~\footnote{According to Resolution 1 of the 26th
    Conference of 
  Weights and Measures~(Comptes Rendus des s{\'e}ances de la
  vingt-sixi{\`e}me Conf{\'e}rence G{\'e}n{\'e}rale des Poids et
  Mesures. R{\'e}solution 1, Annexe 3, BIPM, 2019),
  as of 20 Mai  
  2019
  the
  Boltzmann
  and the
  Planck
  constants have the fixed values
  $\kb=1.380\,649\,10^{-23}\;\chem{J/K}$
  and
  $h=6.626\,070\,15\;10^{-34}\;\chem{Js}$
  , respectively. 
}. 
The MSD increases quadratically with time and
such a behavior is termed \textit{ballistic
  motion}.

In quantum mechanics, the time dependent expectation value of the
particle's position can be given as the trace 
$x(t)=\Tr{x}{(t)}$, where $\hat{x}$ is the position operator and 
$\rhoop(t)$ is the time dependent density operator. 
At thermal equilibrium, the ensemble averaged density operator is
constant,
$\rav{\rhoop(t)}{\theta}=\rav{\rhoop(0)}{\theta}$, so that 
$x(t)=x(0)$ and $\dxsq(t)\equiv 0$. This is equivalent, however, to
the classical result, if $\rav{v}{\theta}$ is taken instead of $v$ 
to calculate 
$\dxsq(t)$. Rather than evaluating the trace with the
ensemble averaged density operator, one should evaluate it with a
typical member of the thermal
ensemble~\cite{Tolman:1938}, and perform the average afterwards. How does
$\dxsq(t)$ then look like? 
To the best of our knowledge, this question
does not seem to 
have received the appropriate attention so far. It is the subject of
the present work.

Diffusion has been understood as the result of stochastic processes 
involving many body interactions since more than 
a century~\cite{Einstein:1905,Smoluchowski:1916,Chandrasekhar:1943,Kubo:1992}.  
Theoretical and experimental developments related to diffusion in 
condensed phases and at surfaces have been continuously 
reviewed from the prospective of both classical and quantum mechanics
in the past 50 
years, and 
references~\citenum{Klinger:1974,Klinger:1983,Haenggi:1990,Gomer:1990,Lombardo:1991,Pollak:1994,Bisquert:2007p,Ellis:2009,Wu:2010,Jonsson:2011,PriceAlonso:2013} 
can only  partially cover the vast literature on this matter. 
 If a particle is undergoing diffusion,
its MSD increases linearly with time, quite contrary to the ballistic
motion of a free particle. This property might not be unique to
diffusion, however. 

Aside this well established consensus,  
 the delocalized nature of the quantum mechanical state of a
 \textit{single} particle  
 inherently reflects the diffusive character of its
 motion~\cite{Schroedinger:1931,Nelson:1966}. F\"urth
 related this character to the uncertainty
 principle~\cite{Fuerth:1933}. 
 The thermal probability density of a single adsorbate, for instance,
 is extremely
 delocalized on the adsorption substrate and the uncertainty to find a
 particle at a specific position in space increases with the size of
 the space. Under these circumstances it
 seems worth to investigate how the temporal 
evolution of the mean square displacement of a particle's position looks
like, when the particle's dynamics is described entirely by quantum 
mechanics, say, from  the solution of the Liouville-von-Neumann 
equation for a thermal state or, equivalently, from the solution
of the Schr\"odinger equation with an initial thermal wave packet.

Because quantum mechanical delocalization is a peculiar property of a
single particle, this   
investigation is conveniently conducted, in a first approach, in the
independent particle 
formalism of a many body system. This will be the approach adopted in
the present work.
Any possible interaction among the particles or
between the particles and their environment that leads to a
dissipation of energy will be discarded. 
Friction, which is related to 
random many body interactions via the fluctuation-dissipation theorem,
is consequently excluded in the independent particle formalism. 
One may argue that, 
under these conditions, it is meaningless to study diffusion.  
While it is conceptually 
interesting to reopen the question of diffusion without
friction~\cite{Fuerth:1933,Wallstrom:1994}, this is not the aim of the
present work. Rather, the focus will be on
properties of the time dependent MSD of an
independent, thermalized quantum particle, on 
their relation to properties of diffusion and, quite
critically, on their potential relation to observables.  
Obviously, the neglect of many body interactions is a strong
approximation and the present approach will correspondingly yield
approximate results, only. 

Scattering
experiments
yield information 
on particle diffusion 
by means of the transformation of the inelastic scattering 
cross section to the 
space-time pair correlation function $G(r,t)$ formulated by van
Hove~\cite{vanHove:1954}.
The space Fourier transformation
of the pair correlation function is the intermediate scattering
function (ISF). 
The space-time Fourier transformation
of the pair correlation function is the dynamical structure factor
(DSF). Diffusion coefficients can be extracted as rate constants from
the decay of the ISF, or as widths of the DSF, depending on the form
of these decays or widths. Many aspects have been covered in the
aforementioned reviews. There are also numerous theoretical
approaches to diffusion that address quantum mechanical effects. None
seem to have addressed the simple question of the quantum dynamical
time evolution of the mean square deviation.

It was pointed out by
Vineyard~\cite{Vineyard:1958} and Schofield~\cite{Schofield:1960} 
that, for sufficiently long times, 
the square of the time dependent width of the Gaussian shaped main peak
($r\sim 0$) of the self-part 
$\subi{G}{s}(r,t)$ of the pair correlation function 
can be interpreted classically as the MSD of scatterers. 
This works particularly well
within the jump diffusion model of Chudley and Elliott in
the continuous diffusion limit of small jumps or long 
times~\cite{ChudleyElliot:1961}.
The quantum mechanical evaluation of the MSD will be discussed 
in relation to this interpretation and to potentially observed
quantities. Of particular interest will be a discussion of results
from helium
scattering experiments of the Xe/Pt(111) system, which mimics an
ideal, two dimensional gas~\cite{Ellis:1999}.

The paper is structured as follows: in section 2 the theoretical
framework of this paper
is explained,
working equations are derived,
and some concepts are defined. This section also contains the main
result of the work, \Eq{dxsqt}, as well as 
the definition of some concepts and of a statistical model to
interpret some of the numerical results.
In section 3 results are discussed. 
Section 4
concludes the work with 
a final discussion of its perspectives.   

\section{Theory}

\subsection{Quantum mechanical approach to $\dxsq(t)$ in the
  independent particle formalism}

In the present work the system to be considered consists of 
an individual particle of mass $m$, the states of which can be described by  
the time dependent density operator $\rhoop(t)$. The latter is
obtained as a solution of the Liouville-von-Neumann equation
$\rhoop(t)=\Uop(t)\rhoop(0)\Uop^\dagger(t)$,
where $\Uop(t)=\exp(-\iu\Hop t/\hbar)$,
$\Hop$ is the system's Hamiltonian, $h$ 
is the Planck constant~\cite{Note1} 
and $\hbar=h/2\pi$.
$\rhoop(0)$ is the initial 
density operator.

A thermal state of a quantum system is appropriately defined in the 
basis of the 
system's eigenstates $\phik{n}$ 
($n=1,2,\ldots$),
whose energies are $E_n$. Without loss of generality, we may assume a
countable set of states. In the ensemble averaged
view of statistical mechanics~\cite{Tolman:1938} 
the henceforth generated 
density operator matrix of a thermal state is diagonal; 
$\bra{\phi_n}\rhotop\ket{\phi_n}\equiv\rhot{nn}$ 
are the time independent thermal populations 
$\rhot{nn}(t)=\exp(-E_n/\kb T)/Q(T)=\rhot{nn}(0)$, while the
off-diagonal elements vanish. Here, $Q(T)$ is the canonical
partition function.
Indeed, in the ensemble average, the thermal density
operator commutes with the system's Hamiltonian and is therefore
constant. In such a view, expectation values of observables in thermal
states are themselves constants. If the MSD of a quantum particle is
defined with $x(t)$ being the expectation value in the thermal state
as given above, this 
quantity vanishes  
identically, independently of whether the particle moves freely or not. 

In 
ref.~\citenum{WojcikDorfman:2003}
the MSD of a quantum particle was defined 
in the context of a uniform quantum
multibaker map 
for 1D quantum walks within a random 
matrix theory. It can be written in terms of a time dependent  evolution as 
$\dxsq(t)={\rm Tr}(\hat{\rho}\;(\hat{x}(t)-\hat{x}(0))^2)$. In this
definition, 
$\hat{x}(t)$ is the time dependent position operator in the
Heisenberg picture and $\hat{\rho}$ is the equilibrium density
operator. We note that there is no operator $\hat{O}$ such that
$\hat{x}(t)-\hat{x}(0)=\Uop^\dagger(t)\hat{O}\Uop(t)$. Therefore, 
the same definition cannot be
given in the Schr\"odinger picture. The result of a quantum dynamical
calculation must not depend on the specific picture used,  
therefore the definition extracted from ref.~\citenum{WojcikDorfman:2003} is 
not further considered here.

In this work we shall view individual members
of the quantum thermal ensemble~\cite{Tolman:1938}.
Such states can be
characterized by a set of random phases  
$0 \le \theta \le 2\pi$, as will be explained below. 
In this view 
the off diagonal elements of the density 
operator matrix are the non-vanishing, time dependent coherences
$\rhot{nm}(t)$ ($n\ne m$).
Coherences 
reflect random fluctuations of the density matrix which do vanish, however, upon   
statistical average. These fluctuations 
are essential for the calculation of the MSD in the present
work. Randomness is also key in a potential approach to link
the quantum mechanical
evolution of the MSD with  
diffusion.  

The mean square displacement (MSD) is hence defined by  
\begin{equation}
  \dxsq(t)=\rav{\left(x(t)-x(0)\right)^2}{\theta}\label{dxsqdef}
\end{equation}
where $x(t)=\Tr{x}{^{(T)}(t)}$ is the quantum mechanical expectation value of the
particle's position at time $t$ in a typical member state of the
thermal ensemble. 
In~\Eq{dxsqdef} the Schr\"odinger picture is used, but the
same quantity is obtained in the Heisenberg picture, as $\Tr{x}{^{(T)}(t)}={\rm
  Tr}(\hat{x}(t)\rhotop)$.

Consistently with the independent particle formalism, coherences are preserved
during the time evolution. The adequateness of this approach to describe the
time evolution of a thermal state in general, and diffusion in
particular, might nevertheless
be questioned. 
The onset of a thermal equilibrium is the consequence of the interaction 
between many particles.
Strictly, to correctly describe the time evolution of the system's
thermal state, this interaction 
must be considered and one has to resort to more
involved open system
quantum dynamical and reduced matrix density
treatments~\cite{Klinger:1983,Kubo:1992,Blum:1996,Vacchini:2009}.
These techniques  
will typically lead to quantum master equations which include
population evolution and decoherence. Steady state solutions of these
equations yield the Boltzmann distribution for the populations of
thermal states. 
Other possibilities are 
path-integral
techniques~\cite{Voth:1996,Manolopoulos:2005,Richardson:2017}, quantum
Langevin or
Bohmian dynamics~\cite{Miret-Artes:2013,Miret-Artes:2018,Miret-Artes:2019}. 

Instead, in the present approach many body interactions are explicitly
excluded. Thermal equilibrium, not its onset, and hence any property
of the particle 
resulting from its contact
with the environment is described by the
imposed initial condition. The dynamical system is closed, which
ensures conservation of level populations and temperature. Any
temporal variation of
the MSD will hence be obtained within closed system equilibrium thermodynamics,
which is a rather uncommon approach to  
diffusion~\cite{Kubo:1992}. Nevertheless, this approach is totally
consistent with 
that sketched in the introductory section for the ballistic behavior
of a free moving 
classical particle. 

\subsection{Specific quantum mechanical expressions for $\dxsq(t)$}

In this section, and for the purpose of a more general use, 
the system considered is a particle of mass
$m$ moving in a 
one-dimensional potential, periodic  
in the lattice constant $a$: 
$V(x)=V(x+a)$.
$\Hop$ is the
corresponding system Hamiltonian:
\begin{equation}
  \Hop = -\frac{\hbar^2}{2\,m}\,\ddxps{x} + V(x)\label{system}
\end{equation}
The system is cast in
periodic super-cells of length $L=N\times a$. The ratio $a/L=1/N$ can
then be considered to be the coverage degree of the one dimensional
lattice.
Later on, results will be discussed for the specific case of a
constant potential, or just $V(x)\equiv 0$. 

Let  
\begin{equation}\label{TWP}
\ket{\psi(0)} = \ket{\psit} \equiv \sum\limits_n\,\frac{\Exp{-\beta\,E_n/2+\iu\,\theta_n}}{\sqrt{Q}}\,\phik{n}
\end{equation}
be an initial \textit{thermal wave packet}  
where the quantities $0\le \theta_n \le 2\pi$ are random
angles, $E_n$ and $\phik{n}$ ($n=1,2,\ldots$) are eigenvalues and
eigenstates of the 
system's Hamiltonian $\Hop$ defined in~\Eq{system}, and $Q$ is the canonical partition function:
\begin{equation}\label{PF}
Q = \sum\limits_{n}\,\Exp{-\beta\,E_n}
\end{equation}
Here and in the following, $\beta\equiv 1/(\kb T)$. The state defined
in~\Eq{TWP} is a typical member of the thermal ensemble~\cite{Tolman:1938,Quack:82b}. 

For $t>0$, the thermal wave packet evolves as the solution of the time dependent Schr\"odinger equation
\begin{eqnarray}
\ket{\psit(t)} &=&
\sum\limits_n\,\frac{\Exp{-\beta\,E_n/2+\iu\,\theta_n-\iu\,E_n\,t}}{\sqrt{Q}}\,\phik{n}
\label{tdTWP}
\end{eqnarray}

The density operator $\rhotop(t)=\ket{\psit(t)}\bra{\psit(t)}$ is the
solution of the 
Liouville-von-Neumann equation with initial $\rhotop(0)=\ket{\psit(0)}\bra{\psit(0)}$. Its matrix elements in the
basis of eigenstates are
\begin{equation}
  \label{rhot}
\rhot{nm}(t)=\bra{\phi_n}\rhotop(t)\ket{\phi_m}=\frac{\Exp{-\beta(E_n+E_m)/2}}{Q}\,\Exp{\iu\,(\theta_n-\theta_m)-\iu\,(E_n-E_m)\,t/\hbar} 
\end{equation}

Eqs.~(\ref{tdTWP}) and~(\ref{rhot}) are valid in an independent
particle formalism and, as assumed throughout this work, in the high
temperature limit, where symmetry restrictions due to the 
indistinguishability of identical particles are not important. 

Following the definition in~\Eq{dxsqdef}, the quantum mechanical MSD
of the particle is then  
\begin{eqnarray}
  \dxsq(t) &=& \rav{\left(\Tr{x}{^{(T)}(t)}-\Tr{x}{^{(T)}(0)}\right)^2}{\theta} =
  \rav{\left(\sum\limits_n\sum\limits_j\,(\rhot{nj}(t)-\rhot{nj}(0))\,x_{nj}\right)^{\ds
      2}\,}{\ds \theta}
\label{dxsqthe}\\
&=&
\frac{1}{Q^2}\;
\sum\limits_{n}
\sum\limits_{j}
\sum\limits_{n\spr}
\sum\limits_{j\spr}
\;
\Exp{-\beta\,(E_n+E_j+E_{n\spr}+E_{j\spr})/2}\nonumber\\
&&
\;\;\;\times\;
x_{n j}
\;
x_{n\spr j\spr}\nonumber\\
&&
\;\;\;\times\;
\left(\Exp{-\iu\,(E_n-E_j)\,t\,/\,\hbar}-1\right)
\;
\left(\Exp{-\iu\,(E_{n\spr}-E_{j\spr})\,t\,/\,\hbar}-1\right)\nonumber\\
&&
\;\;\;\times\;
\rav{\Exp{\iu\,(\theta_n-\theta_j+\theta_{n\spr}-\theta_{j\spr})}}{\ds\theta}
\label{dxsqtheexpanded}
\end{eqnarray}
where $x_{nm} = \bra{\phi_n}\hat{x}\ket{\phi_m}$. 

The averaged quantities $\rav{\Exp{\iu\,(\theta_n-\theta_j+\theta_{n\spr}-\theta_{j\spr})}}{\theta}$
yield zero, unless $n=j$
and $n\spr=j\spr$, or $n=j\spr$ and $j=n\spr$. But for $n=j$ (and
$n\spr=j\spr$) the matrix elements 
$x_{nn}$ (and $x_{n\spr n\spr}$)
vanish, by symmetry, so that only one double sum results in 
the expansion: 
\begin{eqnarray}
\dxsq(t) &=&
  \frac{1}{Q^2}\;
\sum\limits_{n}
\sum\limits_{j}
\Exp{-\beta\,(E_n+E_j)}
\;
\abs{x_{n j}}^2\nonumber\\
&&
\;\;\;\times\;
\left(\Exp{-\iu\,(E_n-E_j)t/\hbar}-1\right)
\,
\left(\Exp{\iu\,(E_n-E_j)t/\hbar}-1\right)
\nonumber\\
&=&
  \frac{4}{Q^2}\;
\sum\limits_{n}
\sum\limits_{j}
\Exp{-\beta\,(E_n+E_j)}
\;
\abs{x_{nm}}^2
\;
\sin^2\left[(E_n-E_j)\,t\,/\,2\hbar\right]
\label{exadxsq}
\end{eqnarray}

Later on it will be shown that, in the case $L\to\infty$ (or zero
coverage degree), this expression yields a simple analytical function
of time.  When $L$ is finite, this expression can be 
evaluated numerically and the numerical results can be rationalized in a model
that simulates hypothetical collisions of particles. 
In that
model it will be assumed that, at some statistically distributed ``collision''
times, the state of the particle changes in such a way that the 
thermal populations are conserved, but the phases 
undergo a complete re-randomization by which coherences are destroyed. 
Suppose now that at a certain time $t$ the particle has undergone a
collision. In 
this case the density matrix has different, uncorrelated phases $\ttheta_n$ at time 0 and
$\theta_n$ at time $t$.
The MSD then results from the average over both sets of random numbers:
\begin{eqnarray}
\dxsq(t) &=&
  \rav{\left(\sum\limits_n\sum\limits_j\,(\rhot{nj}(t)-\rhot{nj}(0))\,x_{nj}\right)^{\ds
      2}\,}{\ds \theta,\ttheta}\nonumber\\
&=&
\frac{1}{Q^2}\;
\sum\limits_{n}
\sum\limits_{j}
\sum\limits_{n\spr}
\sum\limits_{j\spr}
\;
\Exp{-\beta\,(E_n+E_j+E_{n\spr}+E_{j\spr})/2}\nonumber\\
&&
\;\;\;\times\;
x_{n j}
\;
x_{n\spr j\spr}\nonumber\\
&&
\;\;\;\times\;
\left\langle
  \left(\Exp{-\iu\,(E_n-E_j)\,t\,/\,\hbar}\;\Exp{\iu\,(\theta_n-\theta_j)}-\Exp{\iu\,(\ttheta_n-\ttheta_j)}\right)
  \right.\nonumber\\
  &&
  \;\;\;\;\;\;\;\;\;\;
  \left.
\left(\Exp{-\iu\,(E_{n\spr}-E_{j\spr})\,t\,/\,\hbar}\;\Exp{\iu\,(\theta_{n\spr}-\theta_{j\spr})}-\Exp{\iu\,(\ttheta_{n\spr}-\ttheta_{j\spr})}\right)
\right\rangle_{\ds\theta,\ttheta}
\end{eqnarray}
Expansion of the product to be averaged yields
\begin{eqnarray}
  &&
\left\langle
  \left(\Exp{-\iu\,(E_n-E_j)\,t\,/\,\hbar}\;\Exp{\iu\,(\theta_n-\theta_j)}-\Exp{\iu\,(\ttheta_n-\ttheta_j)}\right)
  \right.\nonumber\\
  &&
  \;\;
  \left.
\left(\Exp{-\iu\,(E_{n\spr}-E_{j\spr})\,t\,/\,\hbar}\;\Exp{\iu\,(\theta_{n\spr}-\theta_{j\spr})}-\Exp{\iu\,(\ttheta_{n\spr}-\ttheta_{j\spr})}\right)
\right\rangle_{\ds\theta,\ttheta}\nonumber\\
 &=&
\Exp{-\iu\,(E_n-E_j+E_{n\spr}-E_{j\spr})\,t\,/\,\hbar}\;
\rav{\Exp{\iu\,(\theta_n-\theta_j+\theta_{n\spr}-\theta_{j\spr})}}{\ds\theta}
  \nonumber\\
  &-&
  \Exp{-\iu\,(E_n-E_j)\,t\,/\,\hbar}\;
  \rav{\Exp{\iu\,(\theta_n-\theta_j)}}{\ds\theta}\,\rav{\Exp{\iu(\ttheta_{n\spr}-\ttheta_{j\spr})}}{\ds\ttheta}
  \nonumber\\
  &-&
  \Exp{-\iu\,(E_{n\spr}-E_{j\spr})\,t\,/\,\hbar}\;
  \rav{\Exp{\iu\,(\theta_{n\spr}-\theta_{j\spr})}}{\ds\theta}\,\rav{\Exp{\iu(\ttheta_n-\ttheta_j)}}{\ds\ttheta}
  \nonumber\\
  &+&
  \rav{\Exp{\iu\,(\ttheta_n-\ttheta_j+\ttheta_{n\spr}-\ttheta_{j\spr})}}{\ds\ttheta}
\end{eqnarray}

The averaged factors  
yield zero, unless $n=j$
and $n\spr=j\spr$, or $n=j\spr$ and $j=n\spr$. When these factors do not vanish, the product yields
the value 2. As in~\Eq{dxsqtheexpanded} above, only one double sum results in
the expansion and the MSD becomes thus the time constant quantity
\begin{eqnarray}
\dxsqi &\equiv&
  \frac{2}{Q^2}\;
\sum\limits_{n}
\sum\limits_{j}
\Exp{-\beta\,(E_n+E_j)}
\;
\abs{x_{n j}}^2 
\label{dxsqin}
\end{eqnarray}
which we denote by the MSD symbol $\dxsqi$ marked with a breve.

\subsection{Ideal particles}

In the limit $L\to\infty$, a particle moving in a constant potential
is completely independent and free.  
An independent, free particle will be called \textit{ideal}.  
It is shown in 
appendix~\ref{freeparticle} that the following expression follows
from~\Eq{exadxsq} and 
holds exactly for an ideal, 
thermalized particle of mass $m$:
\begin{equation}
  \dxsq(t) = \frac{\hbar}{m}\,\left(\sqrt{t^2+\tb^2}-\tb\right)\label{dxsqt}
\end{equation}
where 
\begin{equation}
  \tb\equiv\frac{\hbar}{\kb\,T}\label{tb}
\end{equation}
Asymptotically, for times $t\gg\tb$, the MSD becomes a linear 
function of time, $\dxsq(t)\sim2\Dq\,t$, where the quantity 
\begin{equation}
  \Dq=\hbar/2m\label{Dqdef}
\end{equation}
has the dimension of a diffusion coefficient.
\Eq{dxsqt} will be
    further analyzed and discussed in the results section below. 
    It holds strictly for a one-dimensional particle. 
    For the motion on a two dimensional surface 
    or in the three dimensional space, the result on the right hand
    side of~\Eq{dxsqt} is to be multiplied by the
    corresponding dimensionality.

    The time $\tb$ was dubbed \textit{thermal time} in complex time
    Monte Carlo~\cite{Berne:2001} and ring polymer dynamical
    treatments~\cite{Manolopoulos:2007} of quantum time correlation
    functions (see below).  
    
\subsection{Quasi-ideal particles and a statistical model}
\label{qithe}

    A particle moving in a constant potential under periodic boundary
    conditions at finite values of $L$ 
will be called \textit{quasi-ideal}. In appendix~\ref{derdxsqi}, it is 
shown that, 

\begin{enumerate}
  \item 
    for a quasi-ideal, thermalized particle of mass $m$,  
$\dxsqi$ is given analytically by the expression 
    \begin{eqnarray}
  \dxsqi
  &\approx& 
  L\,\hbar\,\sqrt{\frac{2\,\beta}{\pi\,m}}\;J\left(\frac{\hbar^2\,\beta}{2\,m\,L^2}\right)
  \label{dxsqit}
\end{eqnarray}
where the function $J(y)$ is defined as 
\begin{eqnarray}
J(y) &=&
\frac{\sqrt{2}\pi}{12}\,\erf\left(\frac{1}{\sqrt{2y}}\right)
+
\frac{2\,\sqrt{\pi}}{3}\;\sqrt{y}\;\left(\left(1-\Exp{\ds-\frac{1}{2y}}\right)\,y
-
\left(3-\Exp{\ds-\frac{1}{2y}}\right)\,\frac{1}{4}\right)\nonumber\\
\label{Jy}
\end{eqnarray}
\item the following relations hold:
\begin{eqnarray}
  \dxsq(t) &\le& \dxsqi\\
  \lim\limits_{t\to\infty} \dxsq(t) &=& \dxsqi\label{dxsqti}
\end{eqnarray}
      \end{enumerate}

\Eq{dxsqit} holds the better, the larger $L$. Because
$J(0)=\sqrt{2}\pi/12$,
$\lim\limits_{L\to\infty}\,(\dxsqi/L)=\hbar\sqrt{\pi\beta/m}/6$, so that
$\dxsqi$
scales with $L$ in the limit
$L\to\infty$. Because of this asymptotic behavior, \Eq{dxsqti} holds indeed both
for the ideal and the quasi-ideal particle.

\Eq{dxsqti} shows that $\dxsq(t)$ is asymptotically bound under
periodic boundary 
conditions for finite values of the periodic cell length $L$. The
existence of a  
bound may on one hand be viewed as being artificial and 
technically imposed by the boundary conditions. On the other hand, 
$\dxsqi$ is derived under the assumption that decoherence has taken place
during the motion of the free particle at a hypothetical collision
with a neighboring particle.  

The temporal evolution of the MSD obtained from
Eq.~(\ref{exadxsq}) 
for finite values of the periodic cell length $L$
is therefore subject to the interpretation that it bears 
 some signature of 
 decoherence effects, despite the fact that, technically,
 it is entirely coherent. In the following, a model is proposed based
on a physical interpretation of the periodic boundary conditions for
thermalized states, by
which such a  signature of decoherence can be made evident for a 
particle moving in a constant potential.

In the periodic framework used here, 
classical particles will move concertedly in different cells and 
the minimal distance
between any two particles is $L$. The same holds for localized quantum
particles in a 
coherent state and also for delocalized particles in a thermal state
defined by the \textit{same} set 
of random numbers. Such   
particles would never collide, as they are in different cells. 
In a thermal state, forward and backward directions of motion
are described simultaneously and particles have 
zero average velocities. 
Two particles 
in thermal states with \textit{different} sets of random numbers,
may be regarded as residing in neighboring cells and having
momentarily opposite velocities. They may
therefore collide after some time. 
In performing an ensemble average over random
phases in~\Eq{dxsqthe}, one indeed considers particles in thermal states having
different, uncorrelated  phases. While averaging over different sets
of random numbers one is therefore effectively
taking into account such hypothetical collisions. 
For free particles, the collision time is expected to be proportional
to $L/v$, where $v$ is the expectation value of 
the particles' relative speed, which can range between 0 and $\infty$.

On the basis of this interpretation, the following statistical model
is proposed  
for the
MSD of a quasi-ideal particle moving in a periodic cell of length $L$:
a particle of 
speed $v$ moves freely until a certain time $\alpha\,L/v$, where $\alpha$ is an
adjustable positive parameter; during this period, the MSD
is given by the expression given in~\Eq{dxsqt};
at the time $\alpha\,L/v$, a collision
takes place and the MSD suddenly becomes $\dxsqi$;
a continuous expression for the MSD can then be calculated as an average over collision times or velocities.

Velocities of free particles are
distributed according to the Maxwell-Boltzmann distribution (in one dimension):
\begin{equation}
  p(v) = \sqrt{\frac{2}{\pi}}\,\frac{\Exp{-(v/\vm)^2/2}}{\vm}
\end{equation}
Here $\vm$ is defined as
\begin{equation}
  \vm=\sqrt{\langle v^2\rangle}=\sqrt{\frac{\kb\,T}{m}}=\frac{1}{\sqrt{\beta m}}
\end{equation}
and $\langle v^2\rangle = \int_0^\infty\,v^2\,p(v)\,\dif{v}$. 
The analytical expression for the velocity averaged MSD is then
\begin{eqnarray}
  \dxsqa(t)&=&\int\limits_{v=0}^{\alpha\,L\,/\,t}\,p(v)\,\dif{v}\;\dxsq(t)
  +\int\limits_{v=\alpha\,L\,/\,t}^{\infty}\,p(v)\,\dif{v}\;\dxsqi\nonumber\\
  &=&\;\;\;
  \erf\left(\alpha\,\frac{L}{\sqrt{2}\,\vm\,t}\right)\;\vm^2\tb^2\,\left(\sqrt{(t/\tb)^2+1}-1\right)
  \nonumber\\
  &&+\left(1-\erf\left(\alpha\,\frac{L}{\sqrt{2}\,\vm\,t}\right)\right)\;\vm\tb\,L\,\sqrt{2/\pi}\,J\left((\vm\tb/L)^2/2\right)
  \label{dxsqa}
\end{eqnarray}

In the results section, the MSD of a quasi-ideal particle
will be evaluated numerically
from~\Eq{exadxsq}
for variable sizes of the super-cell and  
compared with results from the analytical expression
from~\Eq{dxsqa} to assess the appropriateness of
the model.
In the context of this model, the time
\begin{equation}\label{tc}
  \tc=L\,/\,\vm = L\,\sqrt{\frac{m}{\kb\,T}}
\end{equation}
can  be interpreted as being an average collision time for the otherwise free
particle. Note that $\tc/\tb=L/(\sqrt{2\pi}\,\lambda_T^)$, where
$\lambda_T = \sqrt{\hbar^2\beta/(2\pi m)}$ is the so called thermal de Broglie wave
length~\cite{KittelKroemer:1980,Vacchini:2009}. 
 
\subsection{Pair correlation, ISF and DSF of an ideal particle}

The mean square displacement (MSD) is an observable quantity. 
In scattering experiments, 
actual observables are widths of the dynamical structure factor (DSF)
for quasi-elastic scattering or decay rates of the intermediate
scattering function (ISF).

Van Hove gave the expression for the
self-part of the pair correlation function $\Gs(x,t)$ of an ideal gas
particle~\cite[page 256]{vanHove:1954}:
\begin{eqnarray}
  \Gs(x,t)
  &=&
  \frac{\Exp{\ds -\frac{x^2}{2\,\dxsqc(t)}}}
       {\sqrt{2\pi\;\dxsqc(t)}}\label{Gsfp}
\end{eqnarray}
where
\begin{equation}
  \dxsqc(t) = \vm^2\,t^2 - \iu\,2\,\Dq\,t
\end{equation}
is a complex valued squared length. The quantities $\vm$ and 
$\Dq$ are defined as given above. Formally  
\begin{equation}
  \int\,\dif{x}\;x^2\,\Gs(x,t) = \dxsqc(t)\label{dxsqGs}
\end{equation}
For times $t\gg 2\Dq/\vm^2 = \hbar\beta = \tb$,
this complex number becomes
essentially real, 
$\dxsqc(t) \approx \vm^2\,t^2$ which corresponds to the MSD 
expected for a classical, free moving
particle. This correspondence concomitantly
led to the aforementioned  
interpretation~\cite{Vineyard:1958,ChudleyElliot:1961}.  

The DSF is the space-time Fourier transform of the
pair correlation function which, when reduced to the self-part of such
a gas, yields 
\begin{eqnarray}
  S(q,\om) 
  &=&
  \frac{\Exp{-\frac{(\om-\Dq\,q^2)^2}{2\,\vm^2\,q^2}}}{\sqrt{2\pi\,\vm^2\,q^2}}\;\label{DSFfp}
\end{eqnarray}
where $\hbar\om$ is the energy loss of the scattered beam. The
energy shift $\hbar\Dq q^2 = \hbar^2q^2/(2m)$ is the  the
``recoil energy''~\cite{Schofield:1960}.

The ISF  
 is the space Fourier transform of the pair correlation
function which, in the case of the ideal gas, has 
the simple form
\begin{eqnarray}
  I(q,t) 
    &=& \frac{\Exp{-\frac{\dxsqc(t)\,q^2}{4}}}{\sqrt{2\pi}}\label{ISfp}
\end{eqnarray}
The amplitude of the ISF of an ideal gas evolves with the real part
$\Re\dxsqc(t)$, i.e. following the classical expression of the MSD of an
ideal particle, and decays hence quadratically with time.
The \textit{phase} of the ISF of an ideal gas
\begin{eqnarray}
  \PISF(q,t) 
    &=& \frac{\Dq\,t\,q^2}{2}\label{PISF} 
\end{eqnarray}
evolves with the 
imaginary part $\Im\dxsq(t)=2\Dq\,t$ and is quite obviously related
with the quantum mechanical expression for the MSD of an ideal
particle.

\section{Results and discussion}

\subsection{Ideal particles}

\Eq{dxsqt} is derived in the
appendix and holds for an ideal,
thermalized  particle. Ideal gas
molecules moving at temperature $T$ are ideal   
particles at thermal equilibrium and a 
possible approximate realization of an ideal
 particle is a 
 noble gas at high temperature and low pressure. Another example is
 that of xenon atoms adsorbed on platinum, which behave almost as an ideal,
 two-dimensional gas~\cite{Ellis:1999}. Before addressing
 this example below, the result contained in \Eq{dxsqt} merits some 
 comments.

 First, in the classical limit $\hbar\to 0$, $\dxsq(t)$ vanishes
 identically. As explained above, $\dxsq(t)$ does also vanish identically
 within the ensemble 
 averaged view of the equilibrium density matrix. The
 definition given by~\Eq{dxsqdef} yields therefore a pure quantum mechanical
 quantity. However, for $\hbar> 0$ and initial times
 $0\le t\lesssim\tb = \hbar\beta$, 
 the ideal thermalized particle behaves similarly in classical and
 quantum mechanics, with 
 $\dxsq(t) = (\kb\,T/2m)\,t^2+O(t^4)$ and an effective classical
 temperature that is half the quantum mechanical value.

\begin{figure}[h]
 \begin{center} \includegraphics[width=8cm]{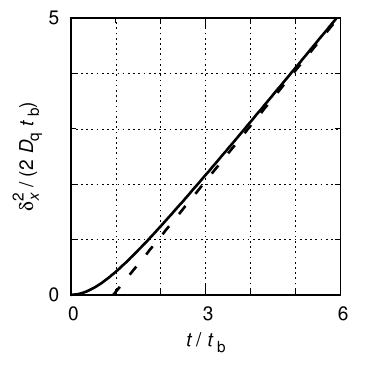}
    \parbox{14cm}{
      \caption{\label{dxsq0} Quantum dynamical time evolution 
        of the
        mean square 
        displacement $\dxsq(t)$ for an ideal particle of mass $m$
        at temperature $T$ according to~\Eq{dxsqt}; for
        $t\lesssim\tb\equiv\hbar/\kb T$, the motion is ballistic;
        for $\tb\ll t$, the motion resembles Brownian diffusion 
        $\dxsq(t)=2\Dq\,t$; the quantity 
        $\Dq$, defined in~\Eq{Dqdef},
        is half the slope of the interrupted line.
}
    }
    \end{center}
\end{figure}

Secondly, the ideal 
quantum particle shows, after some initial time,  a typical feature 
of Brownian diffusion, in that $\dxsq(t)\propto t$.
The characteristic thermal time $\tb$ marks somehow the transition
from ballistic motion 
to Brownian diffusion-like motion (see~\Fig{dxsq0}). 

Considering alone this behavior of the MSD, the ideal quantum particle
seems, thirdly,  to undergo Brownian diffusion even in
the absence of friction. In his seminal work~\cite{Einstein:1905},
Einstein postulated a  dynamical equilibrium
between the motion of a suspended classical particle due to an external force
acting on it and the gas kinetical diffusion process,
which leads to a diffusion coefficient that is proportional to 
temperature and inversely proportional to the friction constant.
A classical particle with zero friction has an infinite diffusion coefficient. 
In contrast, the ideal quantum 
particle seems to have an intrinsic, temperature independent diffusion
coefficient, which can be generally given by the formula  
$\Dq = \lim\limits_{t\to\infty} \dxsq(t)/(2dt) = \hbar/2m$ (\Eq{Dqdef}), where
$d=1$ is the dimensionality of the system. This formula was proposed
by Nelson~\cite{Nelson:1966}, but probably used for the first time by
F\"urth~\cite{Fuerth:1933}. 

Finally, propagating a wave function for a free particle  along the negative
imaginary time axis 
is equivalent to solving Fick's second law for the
wave function with the
diffusion coefficient $\Dq$~\cite{Schroedinger:1931,Fuerth:1933}.
\Eq{dxsqt}, which is related to the evolution of the 
physical observable $\dxsq$ in real time, might therefore not be
unexpected. However, this equation 
cannot obviously be derived from a propagation in imaginary time. Note
that the ensemble averaged density of thermalized ideal or quasi-ideal particles
is a constant, both in space and time, satisfying thus trivially Fick's
law without, however, enabling us to extract a diffusion constant from
this law. 

The quantity $\hbar/m$ has
been reported previously to be related to the ``quantum limit'' of
diffusion. Enss and Haussmann determined a limiting value of
$\sim 1.3\,\hbar/m$
for the spin diffusivity in the unitary Fermi gas using
the strong-coupling Luttinger-Ward theory~\cite{Enss:2012}, for which
experimental evidence was given~\cite{Sommer:2011a,Sommer:2011b}.
Bruun~\cite{Bruun:2011} derives
a variational expression for the spin diffusion coefficient  from the
Boltzmann-Landauer equation, which is proportional to $\hbar/m$ with
dependencies on temperature varying from $T^{- 2}$ to $T^{-1/2}$,
  depending on the coupling and temperature regimes, and determines a
   minimum of $1.1\,\hbar/m$; a quantum limited shear diffusion
   constant of $0.5\,\hbar/m$ was also reported in
   ref.~\cite{Enss:2011}. While in these papers $\hbar/m$ expresses a
   \textit{lower bound} for the 
diffusion constant, 
Shapiro~\cite{Shapiro:2012} 
calls $\hbar/m$ a ``natural unit'' for diffusion in cold-atoms
diffusion
, which ``signals a
crossover to a purely 
quantum  mode of transport'', and Semeghini and co-workers report on
experimental estimations of the ``upper quantum transport limit''
$\hbar/3m$ from the 
measurement of the mobility edge for 3D Anderson
localization~\cite{Semeghini:2015}.
While all these papers deal with more or less strongly and randomly
coupled many particle 
systems, in the present work the expression for $\Dq$ is obtained just from the
random  
fluctuations of the thermal quantum state of an independent particle
in the absence of any 
interactions. To our knowledge, such an 
analytical result is unprecedented.

The diffusion of a particle immersed in a
fermionic~\cite{Ray:2010} bath was analyzed from numerical solutions
of a generalized Langevin  
equation for the quantum mechanical expectation value of the
particle's position. Here, too, the MSD of the particle undergoes a
transition from ballistic 
motion to Brownian diffusion with a temperature dependent 
diffusion constant.
Unbounded diffusion with ballistic and 
Brownian diffusion limiting cases was reported as being a
consequence of spreading wave packets in the framework of Peierls
substitutions for different regimes of the statistics of
eigenstates~\cite{Geisel:1991}. In the present work, no 
randomness is supposed for the values of the eigenstates. 

Mousavi and Miret-Art\'es report on classical and
Bohmian MSD of a free particle subjected to a frictional force in the
spirit of the Ornstein-Uhlenbeck stochastic process 
(equation 78 and figure 2 in
ref.~\citenum{Miret-Artes:2019}). 
As in~\Eq{dxsqt},
the MSD derived in ref.~\citenum{Miret-Artes:2019} changes character from 
ballistic motion to Brownian diffusion. The diffusion coefficient is 
 temperature and friction 
dependent, however,  in agreement with Einstein's
formula. Figure 1 of 
ref.~\citenum{Miret-Artes:2019} shows the time evolution
of the 
uncertainty product: $\Delta x \Delta p$ 
is initially minimal, then increases with time to a maximum and
finally drops to its minimal value again in the long time limit.
In the present study the uncertainty product is 
always maximally infinite, due to the complete
delocalization of the thermal wave packet describing the free
particle.

The stochastic superposition of localized 
Gaussian wave packets used in~ref.~\citenum{Miret-Artes:2019} 
results in the classically expected expression for the MSD of a free particle in
the limit of zero friction. 
In the next section this result is 
analyzed in more detail.

\subsection{Xe/Pt(111) as an ideal particle system}

A system that comes close to the ideal particle is Xe/Pt(111). In
ref.~\citenum{Ellis:1999}, cross sections of this system were 
measured in time-of-flight (TOF) 
experiments with helium atoms scattered at low coverage. The form of the
DSF used for the  
analysis in that work differs from that of~\Eq{DSFfp} in that the
recoil energy was neglected, as is usually done in
classical evaluations of 
the DSF. 
At a typical momentum transfer wave number $q\approx 1\,\proA$
the recoil energy of a xenon atom is
approximately 0.016~meV. It is indeed very small
compared to the thermal energy $\kb T \approx 9$~meV at the surface
temperature of 105~K, and two orders of magnitude smaller than the
width of the DSF (1~meV). 
It is not really measurable at the
precision of the experiment.
Because the DSF from~\Eq{DSFfp}, or its
classical approximation, fits well the measured data, the conclusion from
ref.~\citenum{Ellis:1999} is that xenon moves ballistically as a
nearly ideal, classical  
two-dimensional gas on an essentially flat potential energy surface
with nearly no friction and $\dxsq(t) = \vm^2\,t^2$ for times much
longer  than $\tb \approx 73$~fs at the temperature of the
experiment.

The pair correlation function is not
a directly measurable quantity, however. The squared length $\dxsqc(t)$
is a complex number and 
therefore 
the pair correlation function cannot be the time dependent
expectation value of a Hermitian operator. The same holds for the
intermediate scattering function, which belongs to the class of
quantum time correlation functions~\cite{Kubo:1992,Berne:2001}. 
The interpretation extracted   
from~\Eq{dxsqGs} might not be sufficient to complete our knowledge about the
actual motion of the 
particle, for which 
a direct measurement
of the MSD would be needed. 

The quantum mechanical result of~\Fig{dxsq0} is of course itself
subject to experimental 
verification. One way to verify it 
 would be to use data from 
spin-echo experiments which can resolve the real and imaginary part of the
ISF~\cite{Ellis:2009}. As can be seen from~\Eq{PISF}, 
the long time behavior of $\dxsq(t)$
from~\Eq{dxsqt} could then be determined from $\PISF(q,t)$.  
It would be interesting to determine this function  for the
Xe/Pt(111) 
system which, so far, does not seem to be known experimentally. No
other system similar to Xe/Pt(111) seems to be known, either, 
that would reflect experimentally the conditions imposed to
derive~\Eq{dxsqt}. It should be noted, however, that 
the period $\subi{\tau}{p}$, for which
$\PISF(q,\subi{\tau}{p})=2\pi$, increases quadratically
with the relaxation time of the amplitude of the ISF for an ideal
particle, so that  
the experimental verification of \Eq{PISF} will be challenging. For
Xe/Pt(111), this verification is rendered additionally difficult because the
ISF decays very fast. Finally, as was discussed in
ref.~\citenum{Miret-Artes:2007}, even small friction leads to a 
reshape of the quasi-elastic peak of the DSF, the so-called motional
narrowing effect, so that $\PISF$ might be expected to change
considerably in the presence of friction.  In this context it is worth
mentioning 
that equations 43 and 44 in ref.~\citenum{Miret-Artes:2018} match exactly
Eqs.~(\ref{DSFfp}) and~(\ref{ISfp}) from the present work.

It would likewise be interesting to investigate whether $\PISF$ can more
generally be related to the quantum mechanical expression for the
MSD in situations where the particle is not free. Particularly
suitable would be the investigation of    
systems with a high PES corrugation and low friction, such as
Cs/Cu(100), in order to check for  
potential deviations
from~\Eq{PISF} that could arise from surface corrugation. Information
on the real and imaginary parts of the ISF for this system are
depicted in figure 3 of ref.~\citenum{Ellis:2007}. The consideration
of background count rates and compensation for elastically scattered
beam components inhibit a direct evaluation of $\PISF$,
however, from that work. Additionally, 
the derivation of a potential energy surface for this system  
from \textit{ab initio} calculations is pending. 

An easier  assessment in the context of the present work offers the
CO/Cu(100) system, 
for which a prediction of the MSD can be made from quantum dynamics 
for a reasonable long 
time interval before effects from the neglect of friction are expected
to set in. This work is in progress and the focus in the 
remaining part of this work is therefore devoted to this system.

\subsection{CO molecules as ideal and quasi-ideal particles}

For a zero (or constant) potential $V(x)\equiv 0$,
$E_n=\hbar q_n^2/(2m)$, and the matrix elements $x_{nn\spr}$ are given
analytically (see appendix~\ref{freeparticle}). The sum
in~\Eq{exadxsq} is carried 
out numerically and yields the MSD for what was termed a quasi-ideal
particle above.
The MSD of an ideal particle is given by~\Eq{dxsqt}.

\begin{figure}[h]
  \begin{center}
    \includegraphics[width=10cm]{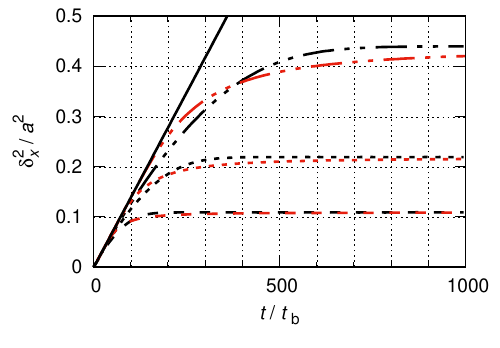}
      \parbox{14cm}{\caption{\label{dxsq-test} Quantum dynamical time evolution 
        of the
        mean square 
        displacement (MSD) $\dxsq(t)$ of a CO molecule ($m\approx
        28$~u) moving on a hypothetically flat Cu(100) 
        surface ($a\approx 256$~pm) at a temperature $T=190$~K.
        Black interrupted lines show the MSD for quasi-ideal particles
        according to~\Eq{exadxsq} for finite
        values of $L$:
        $L=10\,a$
        ($\protect\rule[1mm]{2.5mm}{0.1mm}\;\protect\rule[1mm]{2.5mm}{0.1mm}\;\protect\rule[1mm]{2.5mm}{0.1mm}$), 
        $L=20\,a$ 
($\protect\rule[1mm]{1.1mm}{0.1mm}\;\protect\rule[1mm]{1.1mm}{0.1mm}\;\protect\rule[1mm]{1.1mm}{0.1mm}\;\protect\rule[1mm]{1.1mm}{0.1mm}$),
        $L=40\,a$ 
        ($\protect\rule[1mm]{2.5mm}{0.1mm}\;\protect\rule[1mm]{1.1mm}{0.1mm}\;\protect\rule[1mm]{1.1mm}{0.1mm}\;\protect\rule[1mm]{2.5mm}{0.1mm}$).
        The continuous black line is for $L\to\infty$
        (ideal particle, \Eq{dxsqt}).
The
        red interrupted lines give results from~\Eq{dxsqa} with an optimized
        parameter $\alpha=0.35$. 
        The
        time unit is  
        $\tb = \hbar\beta \approx 40$~fs.
}
    }
    \end{center}
\end{figure}

The lines shown in~\Fig{dxsq-test} give the MSD of CO molecules moving at temperature 190~K as ideal and
quasi-ideal particles of mass $m\approx 28$~u 
on a Cu(100) surface that is completely flat, i.e. where
the potential energy is constant.
Any corrugation or barrier that could hinder the free 
motion is thus entirely removed. 
System and temperature were chosen for the sake of
comparison with the MSD of a
CO molecule moving along the nearest neighbor direction on a perfect Cu(100)
substrate ($a\approx 256$~pm), to be discussed in a separate publication.

The interrupted black lines show the MSD of the quasi-ideal particle moving in 
super-cells of lengths $L=N\times a$, where $N=10$, 20 and 40 (see
caption).
They correspond to a 10,  5 and 
2.5\% coverage degree, respectively. Bases contain 
$K = 100\times N$ functions, i.e. a constant number of functions per elementary
cell of length $a$, 
and results are numerically 
converged.
The continuous black line shows the
 MSD for the ideal particle, corresponding to a  
 0\% coverage degree situation.

Initially,  the MSD of the quasi-ideal and ideal particles match.
The larger $L$, the longer is the agreement.
For very short times of order $\tb$, 
because $\sin((E_n-E_j)t/2\hbar)\approx (E_n-E_j)t/2\hbar$,
$\dxsq(t)\propto t^2$ holds also from the numerical evaluation of~\Eq{exadxsq}.

For $t\to\infty$ the MSD for the quasi-ideal particle 
stagnates asymptotically forming a plateau which is also its upper bound. 
The positions of the plateaus 
increase linearly with the super-cell length.
They furthermore agree perfectly with the values 
$\dxsqi(L=10\,a)\approx 0.11\,a^2$,
$\dxsqi(L=20\,a)\approx 0.22\,a^2$ and
$\dxsqi(L=40\,a)\approx 0.44\,a^2$ from~\Eq{dxsqit}. 
Quite remarkably, the asymptotic MSD is only a fraction of the area
$a^2$ of a primitive 
cell, even for super-cells 40 times larger. Boundary
effects therefore influence the MSD evolution rather
dramatically. 
The low asymptotic bound of the MSD must be related to the 
extremely high degree of delocalization of the
thermal wave packet underlying~\Eq{exadxsq} 
throughout the super-cell, and a corresponding small thermal de Broglie wave
length, as discussed below. The delocalization enhances the boundary
effects. 

The time evolution of the MSD for a quasi-ideal particle shows the
characteristic pattern of a confined diffusion~\cite{Ellis:2009}, or of the
diffusion in a Debye crystal~\cite{ChudleyElliot:1961}. Here, it   
can be understood in terms of the statistical model outlined in 
section~\ref{qithe}, as shown by the red interrupted lines in~\Fig{dxsq-test}
which reproduce 
the function $\dxsqa(t)$ defined by~\Eq{dxsqa}, for different values
of the super-cell length $L$, and capture well the behavior of
$\dxsq(t)$ when the value $\alpha=0.35$ is assumed for all
values of $L$.
The good qualitative agreement between the black and red interrupted lines
supports the physical interpretation underlying the model, in 
that the boundary conditions for thermalized states effectively reflect
collisions and decoherence. 
No quantitative agreement should be expected from this rather simple
interpretation, however.  
Addressing its reality
is much beyond the
scope of the present work. 
The model just underlines that the
independent particle picture adopted here for the evaluation of the MSD
looses its
legitimacy after some time  and that a more realistic simulation of the
evolution of the MSD beyond that time would need 
interactions among particles somehow to be included, which would lead
to much more involved quantum master equations such as in
ref.~\citenum{Vacchini:2009}.

\section{Conclusions}

The mean square displacement (MSD) $\dxsq(t)$ of a particle of mass
$m$ at thermal
equilibrium with its environment is an important quantity in atomic
and molecular physics, 
as well as in transport phenomena. While this quantity emerges
naturally from classical statistical mechanics, the search for a corresponding
quantum mechanical expression has not been carried out sufficiently
far, to the best of our knowledge.
At thermal equilibrium, when the 
ensemble averaged view of the density matrix is adopted, any
observable is a constant of time, and the MSD vanishes 
identically. A different result is obtained, when the view is taken up
that the thermal state of the particle is described by a typical
member of the statistical ensemble.

From this prospective, 
a general expression for the MSD of a
thermalized particle moving independently 
in a one dimensional, periodic
potential $V(x)$ is proposed in \Eq{exadxsq} in terms of the
eigenstates and eigenvalues of the particle in that potential, and
temperature.  
It is a key aspect of 
the formalism that a thermal state of the particle is
described by a typical member of the thermal ensemble, rather than by
its statistical average. As a consequence, the probability density shows
fluctuations which are essential to obtain quantum mechanically a time
dependent MSD.
In the case of a free moving particle in an infinitely large space,
$V\equiv 0$, \Eq{exadxsq} yields the simple analytical formula in
  \Eq{dxsqt} for what 
  was termed the MSD of an \textit{ideal particle}.  

\Eq{dxsqt} reveals an interesting behavior: while initially of
ballistic character, the MSD of the ideal quantum particle of mass $m$ 
moving in an
unrestrained manner at temperature $T$ gains the character of 
a Brownian diffusion after a time of order $\tb \equiv \hbar/(\kb\,T)$. 
The temperature independent slope of the MSD obtained at long times  
is given by $\Dq = \hbar/(2\,d\,m)$, where $d$ is the dimensionality of the
particle.
Although $\Dq$ has the dimension of a diffusion coefficient,
these quantities 
should probably not be related as such 
 in the classical understanding of diffusion, due to the absence of
 friction in the present
theoretical framework. 

Albeit interesting, this result also irritates, as it marks a
stark difference between the classical (ballistic) and quantum
(Brownian diffusion) behavior of the MSD for the ideal, thermalized
particle. Because $\dxsq(t)$ vanishes identically in the limit
$\hbar\to 0$, this quantity should be considered as having no
classical counterpart. 
It was nevertheless shown
analytically, 
how the quantum mechanical MSD of an ideal particle can in principle
be extracted from helium-3 spin echo experiments by the measurement of the
phase $\PISF$ of the intermediate scattering function (ISF). For real particles,
neither the quantum mechanical 
MSD nor the ISF are known analytically. The MSD can be obtained
numerically from~\Eq{exadxsq}. The relation between 
$\PISF$ and the MSD remains hypothetical and worth an experimental
verification.  

In numerical   
evaluations with periodic boundary conditions the MSD for a free
particle initially overlaps with the analytical result 
but then evolves asymptotically into a plateau which defines its upper
bound. The 
longer the periodically repeated cell, the higher is the asymptotic
value. This value equals the value the MSD would have gained if, in the
course of its free evolution, the
particle had undergone 
a collision or any other interaction leading to a decoherence of its
density matrix. A simple model based on statistically distributed collision
times allows one to describe qualitatively the numerical result and to interpret
the onset of the plateau as resulting from a decoherence that is
effectively generated by the boundary conditions.   
To verify the truthfulness of this
interpretation,  real many body
interactions need to be considered in a significantly more involved open
system treatment of the dynamics, which is beyond the scope of the
present work.

The present investigation is planed to be extended to a full
dimensional treatment of the dynamics including interactions between
the particles and 
global potential energy 
surface calculated from \textit{ab initio} calculations, for instance
for the motion of carbon monoxide molecules adsorbed on a copper
substrate~\cite{roma:45}.   
Inspired by previous work, inclusion of substrate
atoms in the dynamics in an explicit way~\cite{Meyer:2007}, in a
hierarchical effective mode approach~\cite{Burghardt:2012}, or in the
form of stochastic operators~\cite{Tremblay:2019bb} will allow us 
to address  
more precisely the role of friction on the specific dynamics of CO on
Cu(100) from first
principle calculations. Potentially, non-adiabatic couplings will also
need to be considered, although for CO on Cu(100) they should play a
minor role. Theoretical results can be expected to become
full quantitative once these additional effects have been considered,
which should ultimately allow us to verify   
 the method and to assess the quality of the potential energy surfaces
 calculated \textit{ab initio}. 
To this end, the present work delivers an
important technical information by showing that the 
quantum dynamical simulation of the long time evolution of a thermal state of
the adsorbate requires large grids corresponding to low coverage
degrees of the substrate.

Accurate theoretical simulations of the diffusion of adsorbates on
substrates are important tools to increase our knowledge of this
process. The present investigation has shown that some essential
properties of diffusion emerge  
naturally and quite realistically in a quantum mechanical description
of the adsorbates' 
dynamics. 
Scanning tunneling microscopy (STM) allows one to measure diffusion 
rates at the single atom 
level~\cite{Bradshaw:1997,Ho:2002,Heinz:2004}. STM would therefore be 
 ideally suitable to measure the mean square displacement of xenon
 atoms on platinum and to discriminate whether they move as classical
 or quantum particles. 
With these
techniques the question arises, however, as to what extent the
STM apparatus  does not itself influence the motion of the adsorbed
species~\cite{Giessibl:2008,Grill:2018}.  It will be interesting to
address this question in the context of a full quantum mechanical
simulation of the STM experiment that includes the motion of the
adsorbates.

\clearpage

\begin{acknowledgments}
  The author thanks Peter Saalfrank for bringing to his attention the
  question about the quantum mechanical time evolution of the MSD; 
  very helpful discussions with him as well as with Salvador
  Miret-Art\'es, J\"orn Manz,
  Fabien Gatti, Jean-Christophe Tremblay and Emmanuel Fromager are 
gratefully  acknowledged. This work benefits from
  the ANR grant under project QDDA. 
\end{acknowledgments}
\newcommand{\Aa}[0]{Aa}
%
\newpage
\appendix
\label{appendix}
\renewcommand{\thesection}{\Alph{section}}
\renewcommand{\theequation}{\thesection\arabic{equation}}
\renewcommand{\thefigure}{\thesection\arabic{figure}}
\renewcommand{\thetable}{\thesection\arabic{table}}
\setcounter{equation}{0}
\setcounter{figure}{0}
\setcounter{table}{0}
\setcounter{section}{0}

\section{Free particle}
\label{freeparticle}

We first consider a super-cell of length $L$ with basis functions
$\phi_n(x) = \braket{x}{\phi_n} = \Exp{-\iu q_n x}/\sqrt{L}$,
where $n\in\mathbb{Z}$, and $q_n= n 2\pi / L = p_n / \hbar$, where $p_n$ is the
momentum of the system 
in state $\ket{\phi_n}$. For a free particle of mass $m$, $V(x)\equiv
0$, and these states are eigenstates of the Hamiltonian of~\Eq{system} with
eigenvalues $E_n=p_n^2/2m=\hbar^2 q_n^2/2m$.

The matrix elements
$x_{nj} = \bra{\phi_n}\hat{x}\ket{\phi_j}$ can be expressed 
 analytically:
\begin{eqnarray}
  x_{nj}
  &=& 
  \iu\,\left(
  \frac{2\,\sin((q_n-q_j)\,L/2)}{L\,(q_n-q_j)^2}
  \;-\;
  \frac{\cos((q_n-q_j)\,L/2)}{q_n-q_j}
\right)\label{me-full}
\end{eqnarray}

Note that $x_{nn}$ exists and yields exactly $x_{n n} = 0$.
Because $q_n= 2\pi\,n\,/\,L$, these matrix elements simplify:
\begin{eqnarray}
  x_{nj}
  &=& \left\{
  \begin{array}{lr}
    0&n=j\\
  \iu\,
  \frac{\ds (-1)^{n-j+1}}{\ds q_n-q_j}&n\ne j
  \end{array}
\right.\label{me-simp}
\end{eqnarray}

Insertion in~\Eq{exadxsq} yields

\begin{eqnarray}
\dxsq(t) &=&
  \frac{4}{Q^2}\;
{\sum\limits_{n=-\infty}^{\infty}}\spr
{\sum\limits_{j=-\infty}^{\infty}}\spr
\Exp{-\beta\,\hbar^2\,(q_n^2+q_j^2)/(2m)}
\;
\frac{\sin^2\left[(q_n^2-q_{j}^2)\,\hbar t\,/\,(4m)\right]}{(q_n-q_j)^2}
\label{avedxsq-fp}
\end{eqnarray}

Here the sums extend from $-\infty$ to $+\infty$, and a state with
energy $E_n$ is
doubly degenerate for $n\ne 0$. In these sums, the combination $n=j$ is
explicitly discarded, which is indicated by the prime symbols.

The sums can be replaced by Riemann sums and, approximately, by the  
integrals:

\begin{eqnarray}
\dxsq(t) 
&\approx& 
\frac{4}{Q^2}\,\frac{L^2}{4\pi^2}\;
{\infint\,}\spr\,\dif{q}\;{\infint\,}\spr\,\dif{q\spr}\;
\Exp{-\beta\hbar^2/(2m)\,(q^2+{q\spr}^2)}\;
\frac{
\sin^2\left[(q^2-{q\spr}^2)\,\hbar  t\,/\,(4m)\right]\;
}{(q-q\spr)^2}\nonumber\\
&=&
\frac{L^2}{2\,\pi^2\,Q^2}\;
\infint\,\dif{u}\;\finfint\,\dif{v}\;
\Exp{-\beta\hbar^2/(2\,m\,L^2)\,(u^2+v^2)}\;
\frac{
  \sin^2[\hbar\,u\,\,v\,t\,/\,(2\,m\,L^2)]
}{v^2}
\nonumber\\
\label{dxsqriemann}
\end{eqnarray}

The prime symbols keep the same signification. In the second equation,
the variable substitutions $u=(q+q\spr)L/\sqrt{2}$ and 
$v=(q-q\spr)L/\sqrt{2}$ were adopted and the integral over $v$ is understood as the principal value (symbol
$\fint$). The replacement of the sums by 
Riemann integrals invariably leads to errors. Their relevance will be discussed in detail below.

The two integrals can be evaluated separately. The integral over $u$
yields

\begin{eqnarray}
F(v,t) 
&=& 
\infint\;
\Exp{-\beta\hbar^2/(2mL^2)\,u^2}\,\sin^2\left(\frac{\hbar\,v\,t}{2\,m\,L^2}\,u\right)\;\dif{u} 
\nonumber\\
&=& 
\sqrt{\frac{\pi\,m}{2\,\beta\,\hbar^2}}\,L\;
\left(1 - 
\Exp{-\frac{\ds t^2}{2\beta\,m\,L^2}\,v^2}\right)
\label{Ffunction}
\end{eqnarray}

Consider the two characteristic times: $\tc = \sqrt{\beta\,m}\,L$ and $\tb=\hbar\beta$. Then the function
$F(v,t)$ can be expressed as 
\begin{eqnarray}
F(v,t) 
&=& 
\sqrt{\frac{\pi}{2}}\,\frac{\tc}{\tb}\;
\left(1 - 
\Exp{-\frac{\ds t^2}{2\,\tc^2}\,v^2}\right)
\end{eqnarray}

Note that, in the limit $L\to\infty$, the following expressions hold
for the partition function:
\begin{eqnarray}
  Q &=& \sum\limits_{n=-\infty}^{\infty}\,\Exp{-\beta\,E_n}
  = \sum\limits_{n=-\infty}^{\infty}\,\Exp{-\beta\,\hbar^2\,q_n^2/(2m)}\nonumber\\
  &\approx&
  \frac{L}{2\pi}\;\infint\;\Exp{-\beta\hbar^2/(2m)\;q^2}\;\dif{q} 
  =  L\;\sqrt{\frac{m}{2\pi\beta\hbar^2}}
  \nonumber\\
  &=& \frac{1}{2\pi}\;\frac{L}{\lambda_T}
  = \frac{1}{\sqrt{2\pi}}\;\frac{\tc}{\tb}
\end{eqnarray}

The MSD is then approximately given by the
integral over $v$: 
\begin{eqnarray}
\dxsq(t) &\approx& 
\frac{L^2}{2\pi^2\,Q^2}\;
\finfint\,\dif{v}\;
\Exp{-\beta\hbar^2/(2mL^2)\,v^2}\,\frac{F(v,t)}{v^2}
\nonumber\\
&\approx&
L^2\,\sqrt{\frac{2}{\pi}}\,\frac{\tb}{\tc}\;
\pfinfint\,\dif{v}\;
\Exp{-\frac{\ds \tb^2}{\ds 2\,\tc^2}\,v^2}\,\frac{
  1 - \Exp{-\frac{\ds t^2}{2\,\tc^2}\,v^2}
  }{\ds v^2}
\end{eqnarray}

In the last equation we used the fact that the integrand is an even
function of $v$.

For $\infty>a\ge0$ and $b>0$, the integral 

\begin{eqnarray}
I(a,b) 
&=& 
\int\limits_0^{\infty}\;
\Exp{-b\,x^2}\,
  \frac{1-\Exp{-a\,x^2}}{x^2}
\;\dif{x}\label{Iint}
\end{eqnarray}

has an integrable singularity at $x=0$, so that the principal value exists. 
An analytical expression for it can be
readily obtained:
\begin{eqnarray}
I(a,b) &=& 
\sqrt{\pi}\,\left(\sqrt{a+b}-\sqrt{b}\right)
\end{eqnarray}

While the principal value  undoubtedly exists for $a=0$,
the result shows that for $a\to\infty$, the integral diverges. The assessment of the error made by
substituting the Riemann sums by integrals presented in the next section gives further insight.  

In terms of $I(a,b)$, with $a=t^2/(2\tc^2)$ and
$b=\tb^2/(2\tc^2)$, the MSD is then expressed as

\begin{eqnarray}
\dxsq(t) &\approx& 
L^2\,\sqrt{\frac{2}{\pi}}\,\frac{\tb}{\tc}\;\sqrt{\pi}\;
\left(
\sqrt{\frac{t^2}{2\,\tc^2}+\frac{\tb^2}{2\,\tc^2}}
-\sqrt{\frac{\tb^2}{2\,\tc^2}}
\right)
=
L^2\;\frac{\tb^2}{\tc^2}\;
\left(
\sqrt{\frac{t^2}{\tb^2}+1}-1
\right)
\nonumber\\
\label{dxsqf}
&=&
\frac{\tb^2}{\beta\,m}\;
\left(
\sqrt{\frac{t^2}{\tb^2}+1}-1
\right)
=
\underbrace{
  \frac{\tb}{\beta\,m}
}_{\ds \hbar/m}
  \;
\left(
\sqrt{t^2+\tb^2}
-\tb
\right)
\end{eqnarray}

This is~\Eq{dxsqt}. In two dimensions, because the total square displacement is the
sum of the square displacements in two orthogonal directions, the result is to
be multiplied by the
factor two. Similarly, in three dimensions, the factor three applies.

Despite the fact that the result is independent of
$L$,~\Eq{dxsqf} is an approximation 
to the sum in~\Eq{avedxsq-fp} (and~\Eq{exadxsq}) that becomes better, the larger
$L$. Nevertheless, the 
error can diverge for $t\to\infty$ and any finite value of $L$, as discussed in
the following. 

\section{Free particle, error estimation and a closed formula for $\dxsqi$}
\label{derdxsqi}
The main cause of error between Eqs.~(\ref{dxsqf}) and~(\ref{avedxsq-fp}) is due to
the evaluation of the integral in~\Eq{Iint}. Let
\begin{equation}
  f(x,a) = \frac{1-\Exp{-a\,x^2}}{x^2}
  \end{equation}
where $x,a\in\mathbb{R}^+$. Let $x_n=n\,\Delta x$, with $n=1,2,\ldots$ and $\Delta x >0$. For sufficiently
small
$\Delta x$, $\int_0^\infty\,f(x,a)\,\dif{x} = \sum_{n=1}^\infty\,f(x_n)\,\Delta x + O(\abs{f\dpr}\,(\Delta x)^3)$.
The function $f\dpr$ has an extremum at $x=0$: $\max(\abs{f\dpr}) = a^2/2$. With $a=(t/\tc)^2/2$,
the error will hence evolve as $t^4(\Delta x)^3$.
This evolution is clearly illustrated in~\Fig{dxsq-test} in the main part of the text.

Yet, \Eq{Iint} suggests that
$I(a\to\infty,b) = \int_0^\infty\,\Exp{-b\,x^2}/x^2\,\dif{x}$
  is related to the asymptotic
value $\dxsq(t\to\infty)$, which is finite for finite values of $L$, as shown in~\Fig{dxsq-test}. In the
following a closed analytical formula is developed for this quantity.

From~\Eq{Ffunction}, $F(v,t\to\infty)=\sqrt{\pi\,m/(2\,\beta\,\hbar^2)}\,L$, which can be obtained by
replacing $\sin^2[\hbar\,u\,v\,t\,/\,(2\,m\,L^2)]\equiv 1/2$ in~\Eq{dxsqriemann} and, consequently,
in~\Eq{avedxsq-fp}. Upon replacement, the resulting sum is exactly the expression for $\dxsqi$
in~\Eq{dxsqin}.

In order to obtain a closed formula for this quantity, we use the definition for the matrix elements of the
position operator given in~\Eq{me-full}, i.e. we consider

\begin{eqnarray}
\dxsqi &=&
  \frac{2}{Q^2}\;
{\sum\limits_{n=-\infty}^{\infty}}\spr
{\sum\limits_{j=-\infty}^{\infty}}\spr
\Exp{-\beta\,\hbar^2\,(q_n^2+q_j^2)/(2m)}
\;
X^2(q_n-q_j)
\end{eqnarray}
where
\begin{eqnarray}
  X(q)
  &=& 
  L\,\left(
  \frac{2\,\sin(L\,q/2)}{L^2\,q^2}
  \;-\;
  \frac{\cos(L\,q/2)}{L\,q}
\right)\label{Xfunc}
\end{eqnarray}
Because  $q_n= 2\pi\,n\,/\,L$, $X(q_n-q_j)^2=1/(q_n-q_j)^2=\abs{x_{nj}}^2$. This
change will not alter the result expected for the sum in~\Eq{dxsqin}.
However, it will change the behavior of the corresponding Riemann integral.
It leads to the replacement of the integral in \Eq{Iint} by $J(b)-J(a+b)$, where 
\begin{eqnarray}
J(y) 
&=& 
2\,\int\limits_0^{\infty}\;
\Exp{-y\,x^2}\,
\;\left(\frac{\sin(x/\sqrt{2})}{x^2}-\frac{\cos(x/\sqrt{2})}{\sqrt{2}\,x}\right)^{\ds 2}
\;\dif{x}\label{Jint}
\end{eqnarray}
This integral can be solved analytically and yields~\Eq{Jy}. 
The \textit{error function} $\erf(x)$ is defined as 
\begin{equation}
  \erf(x) =
  \frac{2}{\sqrt{\pi}}\,\int\limits_{-\infty}^x\,\Exp{-y^2}\;\dif{y} - 1
\end{equation}
Because $\erf(\infty)=1$, $\erf(0)=0$ and
$\lim\limits_{y\to\infty} (1-\exp(-1/(2y)))y = 1/2$
\begin{eqnarray}
J(0) &=& \frac{\sqrt{2}\pi}{12}\\
J(\infty) &=& 0
\end{eqnarray}
Consequently, we may write
\begin{eqnarray}
  \dxsqi
  &\approx& 
L^2\,\sqrt{\frac{2}{\pi}}\,\frac{\tb}{\tc}\;J\left(\frac{1}{2}\left(\frac{\tb}{\tc}\right)^2\right)
\nonumber\\
&=& 
L\,\hbar\,\sqrt{\frac{2\,\beta}{\pi\,m}}\;J\left(\frac{\hbar^2\,\beta}{2\,m\,L^2}\right)
\end{eqnarray}
For $L\to\infty$, this quantity scales with $L$ and not with $L^2$, as could have been expected.

The error made in approximating the Riemann sum by integrals can be assessed via the function 
\begin{equation}
  \tilde{f}(x,a) =
  \left(1-\Exp{-a\,x^2}\right)
  \;\left(\frac{\sin(x/\sqrt{2})}{x^2}-\frac{\cos(x/\sqrt{2})}{\sqrt{2}\,x}\right)^{\ds 2}
\end{equation}
The second derivative of this function can be given as $\tilde{f}\dpr(x,a)=c_1(x,a)\,\Exp{-a\,x^2}+c_2(x)$,
where $c_1$ and $c_2$ are analytical and bound on the real axis (for both $x$ and $a$). In the limit
$a\to\infty$ (corresponding to $t\to\infty$), the error is therefore of order $\abs{c_2(x)}\,(\Delta x)^3$,
which is convergent.

\end{document}